\documentclass[prb,amsmath,twocolumn,superscriptaddress,showpacs]{revtex4}
\usepackage{bm}
\usepackage{amssymb}
\usepackage{graphicx}

\def\b0{{\bf{0}}}
\begin{document}
\title{Decoherence and the retrieval of lost information}
\author{Yin Ye}
\affiliation{State Key Laboratory for Superlattices and Microstructures, Institute of Semiconductors,
Chinese Academy of Sciences, P.O. Box 912, Beijing 100083, China}
\author{Yunshan Cao}
\affiliation{School of Physics, Peking University, Beijing 100871, China}
\author{Xin-Qi Li}
\affiliation{Department of Physics, Beijing Normal University, Beijing 100875, China}
\affiliation{State Key Laboratory for Superlattices and Microstructures, Institute of Semiconductors,
Chinese Academy of Sciences, P.O. Box 912, Beijing 100083, China}
\author{Shmuel Gurvitz}
\email{shmuel.gurvitz@weizmann.ac.il}
\affiliation{Department of Particle Physics and Astrophysics, Weizmann Institute of Science, Rehovot
76100, Israel}

\date{\today}


\begin{abstract}
We show that, contrary to common prejudice, a measurement of an open quantum system can {\em reduce} its decoherence rate. We demonstrate this in an example of indirect measurement of a qubit, where  information regarding its state is hidden in the environment. This information is extracted by a distant device, coupled with the environment only. We also show that the reduction of decoherence generated by this device is accompanied by diminution of the environmental noise in the vicinity of the qubit. An interpretation of these results in terms of quantum interference on large scales is presented.
\end{abstract}

\maketitle

\section{Introduction}
The influence of the environment on a quantum system is an issue of crucial importance in quantum information science. It is mainly attributed to decoherence (dephasing), which transforms asymptotically any initial state of the quantum system into a statistical mixture by tracing out the environmental states in the overall  density matrix  \cite{breuer}. The decoherence rate is a basic quantity characterizing the loss of information stored in the quantum system.

In fact, this information is not totally lost, even in the asymptotic limit---it is hidden in the environment. It can be retrieved by a detector that monitors the environment. Such a detector can be placed far away from the quantum system, so that it does not interact with the system directly. The question we address is whether such an information retrieval could influence the decoherence rate of the quantum system. Beyond its fundamental interest, this question is important for a number of applications, for instance quantum cryptography \cite{gisin}. Indeed, if the decoherence rate is increased by the retrieval of information from the environment, it would betray the fact that the system is being monitored by an outside observer.

It is widely accepted that one cannot make a measurement without perturbing the system \cite{gisin}. It is not clear, however, how to make this statement quantitative. For instance, how is the decoherence rate related to the information gain? Even the main premise, that the measurement {\em always} perturbs the system, is still not proven.

We demonstrate in this paper that, contrary to the common assumption, an indirect measurement of an open quantum system can {\em reduce} its decoherence rate. This implies that one can gain information regarding such a system without adding to the disturbance due to the environment, or even diminishing the environmental effects. This peculiar result is a consequence of a large-scale quantum interference effect that weakens local environmental noise near the quantum system.

These questions of decoherence  and information retrieval from the environment are parts of a more general problem: what constitutes the environment for the purpose of decoherence. Indeed, any device monitoring the environment can be considered a part of the environment and then it must be taken into account in evaluating the decoherence rate. Then the same question arises on the next level, regarding devices coupled to the previous set. This will inevitably lead us to the von Neumann hierarchy problem---a system measured by another system and so forth \cite{neu}---and its relevance to decoherence.

In this paper we do not deal with this fundamental problem in its full complexity, but restrict ourselves to the first ``level.'' We therefore consider a two-state quantum system (qubit) coupled to the environment, which does not measure the qubit, but causes its parameters to fluctuate. Then we introduce a distant device (detector), monitoring the environment and indirectly the qubit. For a consistent analysis of this problem it is necessary to treat it entirely quantum mechanically, as the Schr\"odinger evolution of a many-body system, without any {\em ad hoc} ``classical'' or ``stochastic'' assumptions. One can invent models for the fluctuating environment and the measuring devices that are simple enough to allow full quantum mechanical analysis. Such models, however, have to reflect generic features of real fluctuating environments and devices, for instance as considered in  ``control dephasing'' experiments \cite{agu,kouw,neder1}.
\begin{figure}[h]
\includegraphics[width=5cm]{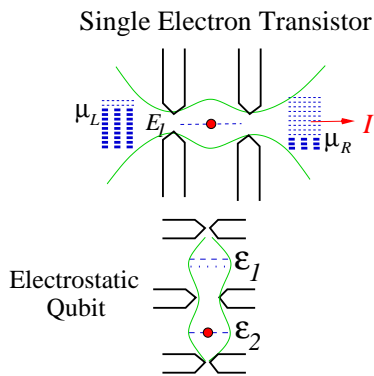}
\caption{(color online) Qubit near a single-electron transistor (SET). Here $\mu_{L,R}$ are
the chemical potentials of the reservoirs. The electric current $I$ through the SET
makes the energy level $\varepsilon_1$ of the qubit fluctuate.} \label{fig1}
\end{figure}

For this reason we represent the fluctuating environment by a single-electron transistor (SET) capacitively coupled to the qubit \cite{gur}, Fig.~\ref{fig1}. Such a setup has been contemplated in numerous solid state quantum computing architectures \cite{kak,makh,goan,gb} and contains most of the generic features of a fluctuating non-equilibrium environment \cite{amnon}. The discreteness of the electron charge creates fluctuations in the electrostatic field near the SET. If the electrostatic qubit is placed near the SET, this fluctuating field affects the qubit by making its energy levels fluctuate, as shown in Fig.~\ref{fig1}. Nonetheless, since the energy level $E_1$ of the SET is deep inside the voltage bias, the SET's current and the fluctuation spectrum of the electric charge inside it are not modulated by the qubit \cite{gur}. Thus such a setup represents the fluctuating (telegraph) noise  acting on the qubit with no back-action \cite{amnon}.
The total current through the SET does not contain any information regarding the qubit state.

The {\em energy resolved\/} current in the reservoir, on the other hand, does carry such information. The energy distribution of the resonant current through the level $E_1$ is given by a Lorentzian centered at $E_1$. If the qubit's electron occupies the level $\varepsilon_1$, the center of the Lorentzian is shifted by the Coulomb repulsion between the two electrons relative to the case when the qubit's electron occupies the lower dot. This implies that by measuring the energy-resolved current through the SET, one can monitor the qubit's state.

\begin{figure}[h]
\includegraphics[width=8.5cm]{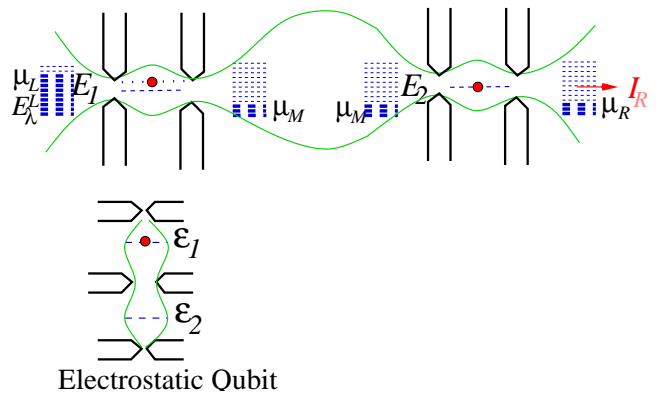}
\caption{(color online) The SET on the right is separated from the one on the left by a middle reservoir. The electric current $I_R$ in the right-hand reservoir (the lead) is sensitive to the state of the
qubit. Here $\mu^{}_{L,M,R}$ are the corresponding chemical potentials.
}
\label{fig2}
\end{figure}

Such an indirect measurement of the qubit  can be performed, for instance, by placing another SET at some distance from the first one, Fig.~\ref{fig2}. In this case the right reservoir (collector) in Fig.~\ref{fig1} becomes the middle reservoir, separating the two SETs in Fig.~\ref{fig2}. As a result, the total current in the right reservoir ($I_R$) becomes dependent on the difference $|E_1-E_2|$: it peaks\cite{g1} when the levels are aligned, $E_1=E_2$.  If the upper dot of the qubit is occupied, the levels $E_{1,2}$ are effectively misaligned due to Coulomb interaction of the qubit with the left SET. Then the current passing the second SET drops, thus indicating that the upper dot of the qubit is occupied. In this way the information on the qubit's state, stored in the energy-resolved current, becomes available.

A similar setup with two distant point-contacts (instead of SETs), coupled with a qubit, has been recently proposed as a sensitive monitor of the qubit's state which can reach high efficiency \cite{kang}. In the present paper, however, we are not concentrating on the efficiency of the qubit's measurement, but rather on the effect of measurement on information stored in the qubit. This information is encoded in the qubit's ``phase'', i.e. in the basis representing a {\em coherent superposition} of the qubit's dot-occupation states. Since the measurement of the qubit via $I_R$, Fig.~\ref{fig2}, reveals its state in the dot-occupation basis, it {\em would\/} affects the qubit's phase (superposition). In what follows we focus on this effect by evaluating the corresponding decoherence rate. A direct measurement of the qubit's phase, is not a subject of this paper.

A detailed quantum mechanical analysis of the entire setup shown in Fig.~\ref{fig2} is presented below. The plan of the paper is as follows. In Sec. II we describe a general framework for treating this system. Then we apply it to the evaluation of the current through separated SETs and the average charge inside the SETs. In Sec. III we evaluate the qubit's decoherence rate and its connection to the current measurement. Sec. IV presents an interpretation and discussion of the results. The last section  is a summary.

\section{Two separated quantum dots attached to a qubit}
\subsection{Description of the model.}

Consider the setup shown in Fig.~\ref{fig2}. The overall Hamiltonian can
can be written as $H=H_d+H_q+H_{int}$, where $H_d$, $H_q$, and $H_{int}$  describe the detector (two separated SETs), the qubit, and their interaction, respectively,
\begin{align}
&H_d=\sum_\lambda \big (E_\lambda^L\,c_\lambda^{L\,\dagger}
c_\lambda^L+E_\lambda^M\,c_\lambda^{M\,\dagger} c_\lambda^M
+E_\lambda^R\, c_\lambda^{R\,\dagger} c_\lambda^R\big)\nonumber\\
&+E_1^{}c_1^\dagger c_1^{}+E_2^{}c_2^\dagger c_2^{}+\sum_\lambda\big
(\Omega_\lambda^Lc^\dagger_1c_\lambda^L
+\Omega_\lambda^Mc^\dagger_1c_\lambda^M\nonumber\\
&+\bar\Omega_\lambda^Mc^\dagger_2c_\lambda^M+\Omega_\lambda^Rc^\dagger_2
c_\lambda^R+H.c.\big)\label{a1}\\[5pt]
&H_q=\varepsilon_1^{}a_1^\dagger a_1^{}+\varepsilon_2^{}a_2^\dagger
a_2^{}+\Omega(a_1^\dagger a_2^{}
+a_2^\dagger a_1^{})\, ,\label{a2}\\[5pt]
&H_{int}=Ua_1^\dagger a_1^{}c_1^\dagger c_1\, . \label{a3}
\end{align}
Here $c_{1,2}^\dagger(c_{1,2}^{})$ and $c_{\lambda}^{L,M,R\, \dagger}(c_{\lambda}^{L,M,R})$ are the creation (annihilation) operators for the electron inside the SETs and in the reservoir ($L$-left, $M$-middle and $R$-right), and $a_{1,2}^\dagger(a_{1,2}^{})$ are the same for the qubit. $\Omega$ is the coupling between the states $a^\dagger_{1,2}|0\rangle$ of the qubit, and $\Omega_\lambda^{L,M,R},\bar\Omega^M_\lambda$ are the couplings between the SET quantum dots and the reservoirs. In the absence of a magnetic field, these couplings are real, but they can be of opposite sign, depending on the relative parity of the states in the quantum dots \cite{lap}. We assume weak energy dependence of all couplings with the reservoirs, $\Omega_\lambda^{L,M,R} =\Omega_{L,M,R}^{}$. For simplicity we consider electrons as spinless fermions.

Consider the dynamics of this system for the case of large bias. This corresponds to the initial condition in which all states in the left lead are occupied and all states in the middle and right leads are empty. In this case one can transform the many-body Schr\"odinger equation to a Lindblad-type master equation for the reduced density matrix, $\sigma (t)$, by tracing out the reservoir states. In our case these are the states of the left, middle and right reservoirs. The reduced density matrix $\sigma_{jj'}(t)$ describes therefore a combined system of the qubit and two SETs. These equations can be written in a general form as \cite{gp,g1,gur}
\begin{align}
&\dot\sigma^{}_{jj^{\prime}}=i({\cal E}^{}_{j^{\prime}}-{\cal E}^{}_j)\sigma^{}_{jj^{%
\prime}} + i \sum_{k}\left(\sigma^{}_{jk}\widetilde\Omega_{k\to j^{\prime}}
-\widetilde\Omega_{j\to k} \sigma^{}_{kj^{\prime}}\right )  \notag \\
&-{\sum_{k,k^{\prime}}}\mathcal{P}_2
\pi\rho(\sigma^{}_{jk}\Omega_{k\to k^{\prime}}\Omega_{k^{\prime}\to
j^{\prime}} +\sigma^{}_{kj^{\prime}}\Omega_{k\to
k^{\prime}}\Omega_{k^{\prime}\to j})
\notag \\
&+\sum_{k,k^{\prime}}\mathcal{P}_2 2\pi\rho\,\Omega_{k\to j}\Omega_{k^{\prime}\to
j^{\prime}}\,\sigma^{}_{kk^{\prime}}\, ,  \label{bp17}
\end{align}
where the indices $j,j',k,k'$ denotes all {\em discrete} states of the system and ${\cal E}_{j}^{}$ is the total energy of the state $j$.
For instance, ${\cal E}=E_1+E_2+\varepsilon_1^{}+U$ for the state shown in Fig.~\ref{fig2}. The second term in the r.h.s. corresponds to {\em direct} hopping between these discrete states. In our case this can take place only between the qubit's states, so that $\widetilde\Omega\equiv\Omega$ in Eq.~(\ref{a2}). The last two terms in Eq.~(\ref{bp17}) represent transitions among the discrete states through the continuum (the reservoir), which has the density of states $\rho_{L,M,R}^{}$. The corresponding hopping amplitudes are therefore $\Omega_{k\to k'}\equiv \Omega_\lambda^{L,M,R}$, Eq.~(\ref{a1}). The first of these terms arises from ``loss'' processes and the second  from ``gain'' processes (borrowing the terminology of the classical Boltzmann equation). $\mathcal{P}_2$ is the Pauli factor: $\mathcal{P}_2=-1$ in  transitions involving two electrons, +1 otherwise. The detailed microscopic derivation of Eq.~(\ref{bp17}) can be found in Ref.~[\onlinecite{gp,g1,gur}]. Here we  only  mention that Eq.~(\ref{bp17}) is valid for any values of coupling and interaction energies, providing that these are much smaller than the applied voltage (large bias limit).

Let us rewrite Eq.~(\ref{bp17}) explicitly for our system of two separated SETs coupled to the qubit, Fig.~\ref{fig2}. The entire system can be found in eight discrete states, denoted as $\{a,a',b,b',c,c',d,d'\}$. Four of them are shown in Fig.~\ref{fig3}. The remaining states, $\{ a',b',c',d'\}$, correspond to those in Fig.~\ref{fig3} with the lower dot of the qubit occupied. Note that transitions between the left and the right SET can proceed only through the continuum, described by the third and the fourth terms in the r.h.s. of Eq.~(\ref{bp17}), whereas transitions between the qubit's states are given by the second term of Eq.~(\ref{bp17}).

\subsection{Resonant current through two separated SETs}

In order to simplify the problem but without losing the main physical features of decoherence, we consider $\Omega =0$ in Eq.~(\ref{a2}). This corresponds to the so-called ``pure decoherence'' model, since no relaxation of the qubit can take place
when $\Omega =0$. This model is widely used in the literature \cite{clive}, and is applicable to a real charge qubit during the most of its operation \cite{def}. In effect, the qubit's states are now decoupled. In this case Eq.~(\ref{bp17}) reduces to three decoupled sets of  equations.
One set couples matrix elements $\sigma_{ij}$ where the states $i,j\in\{a,b,c,d\}$, that is, where the qubit is in its upper state.
The second set couples the $\sigma_{ij}$ where $i,j\in\{a',b',c',d'\}$,
that is, where the qubit is in its lower state.
The third set couples the $\sigma_{ij}$ for $i\in\{a,b,c,d\}$, $j\in\{a',b',c',d'\}$.

\begin{figure}[h]
\includegraphics[width=8.5cm]{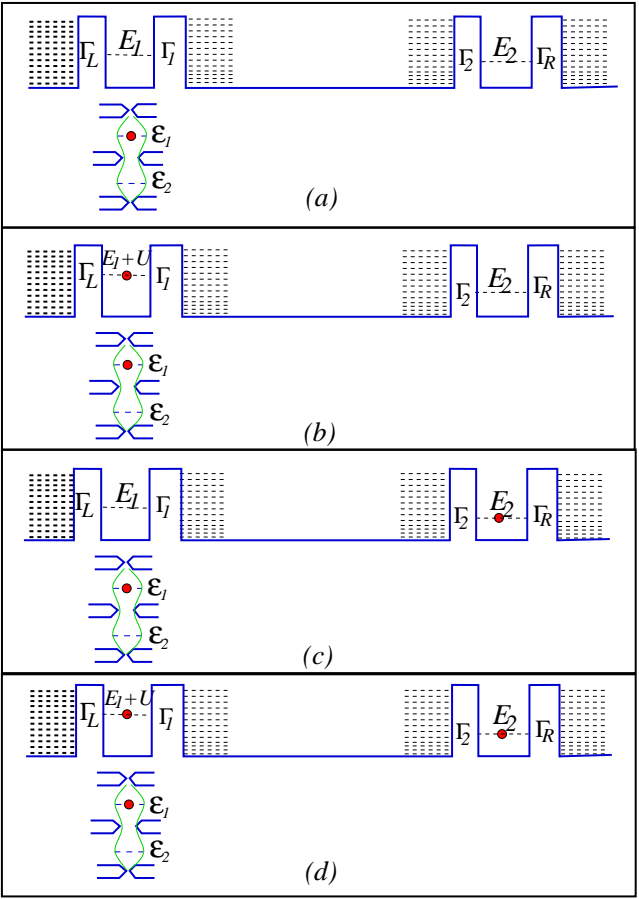}
\caption{(color online) The discrete states of the system with the upper dot of the qubit occupied. } \label{fig3}
\end{figure}

We start with a configuration in which the qubit electron occupies the upper dot of the qubit, as in Fig.~\ref{fig3}. In this case the effect of the qubit on the detector is a permanent energy shift, $E_1\to E_1+U$, due to the Coulomb repulsion between two electrons occupying the qubit and the left SET. One finds from Eq.~(\ref{bp17}) that the corresponding reduced density matrix evolves according to the following equations\cite{fn1},
\begin{subequations}
\label{a4}
\begin{align}
\dot\sigma_{aa}^{}&=-\Gamma_L\sigma_{aa}^{}
+\Gamma_1\sigma_{bb}^{}
+(\Gamma_2+\Gamma_R)\sigma_{cc}^{}+2\Gamma_{12}\,{\rm Re}\,\sigma_{bc}^{}
\label{a4a}\\
\dot\sigma_{bb}^{}&=\Gamma_L\sigma_{aa}^{}
-\Gamma_1\sigma_{bb}^{}
+(\Gamma_2+\Gamma_R)\sigma_{dd}^{}-\Gamma_{12}\,{\rm Re}\,\sigma_{bc}^{}
\label{a4b}\\
\dot\sigma_{cc}^{}&=-(\Gamma_L+\Gamma_2+\Gamma_R)\sigma_{cc}^{}
+\Gamma_1\sigma_{dd}^{}-\Gamma_{12}\,{\rm Re}\,\sigma_{bc}^{}
\label{a4c}\\
\dot\sigma_{dd}^{}&=-(\Gamma_1+\Gamma_2+\Gamma_R)\sigma_{dd}^{}
+\Gamma_L\sigma_{cc}^{}
\label{a4d}\\
\dot\sigma_{bc}^{}&=\left[i(\Delta -U) -{\Gamma_T\over2}\right]\sigma_{bc}^{}-\Gamma_{12}\sigma_{dd}
-{\Gamma_{12}\over2}(\sigma_{bb}+\sigma_{cc}),
\label{a4e}
\end{align}
\end{subequations}
where $\Delta =E_2-E_1$. Here $\Gamma_{L(R)}=2\pi\Omega^2_{L(R)}\rho_{L(R)}^{}$, $\Gamma_1=2\pi\Omega^2_{M}\rho_{M}^{}$, and $\Gamma_2=2\pi\bar\Omega^2_{M}\rho_{M}^{}$ are the partial widths of the levels $E_{1,2}$ due to coupling with the left, right and middle reservoirs, whereas $\rho_{L,R,M}^{}$, are the corresponding density of states. We also denote $\Gamma_T=\Gamma_L+\Gamma_1+\Gamma_2+\Gamma_R$ and $\Gamma_{12}=2\pi\Omega_M\bar\Omega_M\rho_M^{}=\eta\sqrt{\Gamma_1\Gamma_2}$, where $\eta =(\Omega_M\bar\Omega_M)/|\Omega_M\bar\Omega_M|=\pm 1$ corresponds to the relative parity (number of nodes) of the SET states \cite{lap}.

Solving these equations we find the probability of occupation of each SET, $P_{1,2}^{}(t)=\langle\Psi (t)|c_{1,2}^\dagger c_{1,2}^{}|\Psi (t)\rangle$, and the average current in the middle and the right reservoirs, $ I_{M(R)}^{}(t)=\langle\Psi (t)|i[H,\hat Q_{M(R)}]|\Psi (t)\rangle$, where $\hat Q_M=\sum_\lambda c_\lambda^{M\,\dagger} c_\lambda^M$ and $\hat
Q_R=\sum_\lambda c_\lambda^{R\,\dagger} c_\lambda^R$ are the operators for charge accumulated in the middle and the right reservoirs, and $|\Psi (t)\rangle$ is the overall many-body wave function. One finds \cite{g1,gp,gur}
\begin{align}
P_1(t)=\sigma_{bb}^{}(t)+\sigma_{dd}^{}(t),~~
P_2(t)=\sigma_{cc}^{}(t)+\sigma_{dd}^{}(t)
\, ,\label{occ}
\end{align}
and
\begin{align}
I_M^{}(t)&=\Gamma_1^{}P_1(t)+
\Gamma_2^{}P_2(t)+2\Gamma_{12}^{}{\rm Re}\,[\sigma_{bc}^{}(t)]
\nonumber\\[5pt]
I_R^{}(t)&=\Gamma_R^{}P_2(t)
\, .\label{curr}
\end{align}
In a similar way one finds the current in the left lead is given by $I_L^{}(t)=\Gamma_L^{}[\sigma_{aa}^{}(t)+\sigma_{cc}^{}(t)]$.

In the asymptotic limit, $t\to\infty$, Eqs.~(\ref{a4}) can be easily solved by taking into account that $\dot\sigma_{jj'}^{}(t\to\infty )=0$ and using the probability conservation condition, $\sigma_{aa}+\sigma_{bb}+\sigma_{cc}+\sigma_{dd}=1$. Consider the case of $\Gamma_L=\Gamma_1=\Gamma$, corresponding to maximal penetrability of the left dot. Then one obtains for the steady-state currents, $\bar I_{M,R}^{}= I_{M,R}^{}(t\to\infty )$, the following simple expressions,
\begin{subequations}
\label{curr1}
\begin{align}
\bar I_M^{}(U)&=\frac{\Gamma(\Gamma_2+\Gamma_R)[4 (\Delta-U)^2+\Gamma_T(\Gamma+\Gamma_R)]}{8 (\Delta-U)^2 (\Gamma
_2+\Gamma _R)+\Gamma_T^2 \left(\Gamma _2+2 \Gamma _R\right)}
\label{curr1a}\\[5pt]
\bar I_R^{}(U)&=\frac{\Gamma  \Gamma _2 \Gamma _R \Gamma _T}{8 (\Delta-U) ^2
   \left(\Gamma _2+\Gamma _R\right)+ \Gamma _T^2\left(\Gamma _2+2
   \Gamma _R\right)}
\, .\label{curr1b}
\end{align}
\end{subequations}
By evaluating in the same way $\bar I_L^{}=I_{L}^{}(t\to\infty )$, one can easy verify current conservation: $\bar I_L^{}=\bar I_M^{}+\bar I_R^{}$.

\begin{figure}[h]
\includegraphics[width=9.5cm]{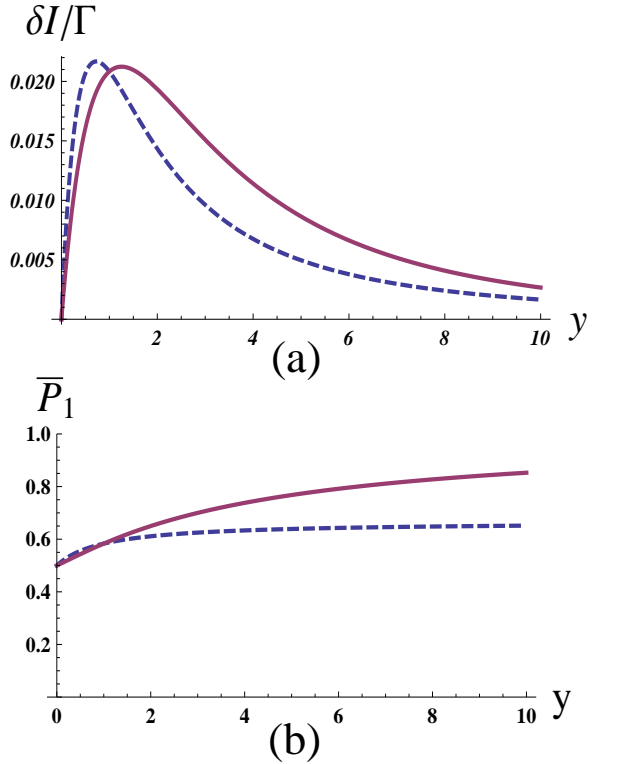}
\caption{(color online) (a) ``Signal'' $\delta I$ and (b) occupation $\bar P_1$ of the left SET as a function of $y=\Gamma_2/\Gamma$ for $\Gamma_L=\Gamma_1=\Gamma$, $\Gamma_R=\Gamma_2$ (dashed line) and
$\Gamma_L=\Gamma_1=\Gamma_R=\Gamma$ (solid line).} \label{fig4}
\end{figure}

If the qubit state corresponds to the lower dot occupied, the currents are given by the same Eqs.~(\ref{curr1}) with $U=0$. Therefore by measuring the change in the current, for instance in the right lead, $\delta I=|\bar I_R^{}(U)-\bar I_R^{}(0)|$, one can determine the qubit state. An example of such a signal for $U=\Gamma$ and $\Delta =0$ is shown in Fig.~\ref{fig4}a. We plot $\delta I$ as a function of the asymmetry between the dots, measured by $y=\Gamma_2^{}/\Gamma$.  The solid line corresponds to $\Gamma_R=\Gamma$ and  the dashed line to $\Gamma_R=\Gamma_2$. The latter corresponds to maximal penetrability of the right SET.

Note that $y=0$ corresponds to switching off the right SET ($\Gamma_2=0$). In this case the current in the right reservoir vanishes and the total current in the middle reservoir is $\bar I_M^{}=\Gamma /2$, see Eqs.~(\ref{curr1}). Therefore the currents become independent of the qubit state and the signal vanishes, as shown in Fig.~\ref{fig4}a. With increased $y$ the signal increases reaching its maximal value near $y=1$, where the penetrability of the system is maximal. Beyond this point the signal decreases with $y$. This is to be expected since for $\Gamma_2\gg U$ the influence of the level displacement $U$ on the current is small.

The probability of finding the left SET occupied in the steady-state limit, $\bar P_1=P_1(t\to\infty )$ Eq.~(\ref{occ}), is shown in Fig.~\ref{fig4}b as a function of $y$ for $\Delta =0$. We find that the presence of the distant second SET increases the occupation probability of the left SET, compared to the case where the second SET is switched off ($y=0$).

\section{Decoherence of the qubit}
Let us assume that the qubit is initially in a linear superposition of its basis states,
\begin{align}
|\psi_0\rangle =\cos\alpha\, a_1^\dagger |0\rangle +e^{i\beta}\sin\alpha\, a_2^\dagger |0\rangle.
\label{inqb}
\end{align}
The qubit's behavior is described by the reduced density matrix, $\sigma_{jj'}^q(t)$, obtained by tracing over the detector states.
The diagonal matrix elements---the probabilities---are given by
\begin{align}
\sigma_{11}^{q}(t)&=\sigma_{aa}^{}(t)+\sigma_{bb}^{}(t)+\sigma_{cc}^{}(t)
+\sigma_{dd}^{}(t)
\label{a5a}
\end{align}
and $\sigma_{22}^{q}(t)=1-\sigma_{11}^{q}(t)$; the initial conditions correspond to Eq.~(\ref{inqb}). Since the qubit's states are decoupled, the probabilities are constant in time, $\sigma_{11}^{q}(t)=\cos^2\alpha$,  $\sigma_{22}^{q}(t)=\sin^2\alpha$ and therefore are not affected by the detector.

This is not the case, however, with respect to the off-diagonal elements of the density matrix,
\begin{align}
\sigma_{12}^{q}(t)&=\sigma_{aa'}^{}(t)+\sigma_{bb'}^{}(t)+\sigma_{cc'}^{}(t)
+\sigma_{dd'}^{}(t)\, .\label{a5c}
\end{align}
This element vanishes at $t\to\infty$ due to the charge fluctuations in the left SET. In order to evaluate $\sigma_{12}^{q}(t)$ and the corresponding decoherence rate, we have to solve the master equation (\ref{bp17}) for the off-diagonal density matrix elements. In the case of
no coupling between the qubit states, $\Omega=0$, these equations can be written explicitly as
\begin{subequations}
\label{a15}
\begin{align}
\dot\sigma_{aa'}^{}&=
-\Gamma_L\,\sigma_{aa'}^{}+\Gamma_1\sigma_{bb'}
+(\Gamma_2+\Gamma_R)\,\sigma_{cc'}\nonumber\\
&~~~~~~~~~~~~~~~~~~~~~~~~~~~~~~~
+\Gamma_{12}(\sigma_{bc'}^{}+\sigma_{cb'}^{})\label{a15a}\\
\dot\sigma_{bb'}^{}&=-(iU
+\Gamma_1)\sigma_{bb'}^{}
-{\Gamma_{12}\over2}(\sigma_{bc'}^{}+\sigma_{cb'}^{})
+\Gamma_L\sigma_{aa'}\nonumber\\
&~~~~~~~~~~~~~~~~~~~~~~~~~~~~~~~~~~
+(\Gamma_2+\Gamma_R)\sigma_{dd'}\label{a15b}\\
\dot\sigma_{cc'}^{}&=-(
\Gamma_L+\Gamma_2+\Gamma_R)\sigma_{cc'}^{}
-{\Gamma_{12}\over2}(\sigma_{bc'}^{}+\sigma_{cb'}^{})\nonumber\\
&~~~~~~~~~~~~~~~~~~~~~~~~~~~~~~~~~~~~~~~~~~~~~
+\Gamma_1\sigma_{dd'}\label{a15c}\\
\dot\sigma_{dd'}^{}&=-(iU
+\Gamma_1+\Gamma_2+\Gamma_R)\sigma_{dd'}^{}
+\Gamma_L\sigma_{cc'}
\label{a15d}\\
\dot\sigma_{bc'}^{}&=i \left(\Delta -U
-i{\Gamma_T\over2}\right)\sigma_{bc'}^{}
-{\Gamma_{12}\over2}(\sigma_{bb'}+\sigma_{cc'})\nonumber\\
&~~~~~~~~~~~~~~~~~~~~~~~~~~~~~~~~~~~~~~~~~~~~~
-\Gamma_{12}\sigma_{dd'}\label{a15e}\\
\dot\sigma_{cb'}^{}&=-\left(i\Delta +{\Gamma_T\over2}\right)\sigma_{cb'}^{}
-{\Gamma_{12}\over2}(\sigma_{cc'}+\sigma_{bb'})
-\Gamma_{12}\sigma_{dd'}\label{a15f}
\end{align}
\end{subequations}
Since the qubit states are decoupled, we have eliminated their energies ($\varepsilon_{1,2}^{}$) from Eqs.~(\ref{a15}) by a gauge transformation.

In the steady-state limit $\dot\sigma_{jj'}^{}(t\to\infty )=0$,  Eqs.~(\ref{a15}) become a system of homogeneous algebraic equations. One can check that in this limit the off-diagonal elements of the density vanishes, so that $\sigma_{12}^q(t\to\infty)=0$, as expected. In order to find the qubit's decoherence rate it is useful to rewrite Eqs.~(\ref{a15}) in a matrix form,
\begin{align}
\dot X(t)+B X(t)=0\, , \label{a8}
\end{align}
where $X(t)$ is a 6-component vector,
\begin{align}
&X\equiv
\{X_k\}=\{\sigma_{aa'},\sigma_{bb'},\sigma_{cc'},\sigma_{dd'},\sigma_{bc'},
\sigma_{cb'}\}\, , \label{a9}
\end{align}
and $B$ is a $6\times6$ matrix (super-operator), corresponding to the
r.h.s. of Eqs.~(\ref{a15}). Note that for a pure-dephasing
model ($\Omega =0$), considered here, the steady state limit depends on the initial qubit's state \cite{clive}.

To solve Eq.~(\ref{a8}) we apply a Laplace transform,
\begin{align}
\tilde X(E)=\int_0^\infty X(t)e^{iEt}dt \, ,
\label{a10}
\end{align}
where $E$ has a small, positive imaginary part.
Then Eq.~(\ref{a8}) becomes an algebraic equation,
\begin{align} (B-i E\, I)\tilde X(E)= X(0)\, , \label{a11}
\end{align}
where $X(0)$ corresponds to the initial state and $I$ is the identity matrix. Solving Eq.(\ref{a11}) we find
\begin{align}
\tilde X_k(E)={\bar D_k(E)\over\det (B-i E I)}\, ,  \label{a12}
\end{align}
where $\bar D_k(E)$ is a corresponding combination of the co-factors
and minor determinants.

In order to find $X(t)$ we have to perform the inverse Laplace
transform
\begin{align}
X_k(t)={1\over 2\pi i}\int_{-\infty}^\infty\tilde X_k(E)e^{-iEt}dE\,
,\label{a13}
\end{align}
where the contour of integration is
$E+i\delta$ with $\delta\to 0$.
We close the contour in the lower half plane.
It follows from Eq.~(\ref{a13}) that
any pole of the integrand at $E=E^{(l)}-i\Gamma^{(l)}$,
obtained from the equation
\begin{align}
D(E)=\det (B-i E I)=0 \label{a14}\, ,\end{align}
results in exponential damping of the off-diagonal density-matrix elements,  $X(t)\propto\exp (-\Gamma^{(l)} t)$. The decoherence rate is defined as $\Gamma_d =\min \{\Gamma^{(l)}\}$, since it dominates the the decoherence at sufficiently large times.

Consider first $\Gamma_2=0$. In this case  the right-hand SET in Fig.~\ref{fig3} is switched off, so that no measurement of the qubit takes place. As in the Sec.~II we consider a symmetric left-hand SET, $\Gamma_L=\Gamma_1=\Gamma$, so that the current flowing through this SET is $\bar I_M=\Gamma/2$. Equation~(\ref{a14}) can be easily solved analytically, thus obtaining for the  decoherence rate of the qubit
\begin{align}
\Gamma_d=\Gamma\left(1-\sqrt{1-{U^2\over 4\Gamma^2}}\right)\, .\label{dec1}
\end{align}
To leading order in $(U/\Gamma)^2$ this is \cite{amnon}
\begin{align}
\Gamma_d\simeq{U^2\over 8\Gamma}={U^2\over2}S_Q(0)\, ,
\label{dec2}
\end{align}
where $S_Q(0)$ is the zero-frequency limit of the power spectrum of fluctuating charge inside the left SET \cite{xin1}
\begin{align}
S_Q(\omega )=2\, {\rm Re}\,\int_0^\infty \langle\delta \hat Q(0)\delta \hat
Q(t)\rangle e^{i\omega t}dt\, . \label{ac1}
\end{align}
Here $\delta \hat Q(t)=c_1^\dagger (t)c_1(t)-\bar P_1$ and $\bar P_1=P_1(t\to\infty )$ is the average charge inside the SET, Eq.~(\ref{occ}). Equation~(\ref{dec2}) shows that dephasing rate is directly related to the noise spectrum of the environment. This result can been obtained by using different methods. In particular, it was proven by Emary \cite{clive} that in the case of pure dephasing ($\Omega =0$), Eq.~(\ref{dec2}) is valid for any type of non-equilibrium environment in the weak coupling limit (here $U^2/\Gamma^2\ll 1$).

\begin{figure}[h]
\includegraphics[width=9.5cm]{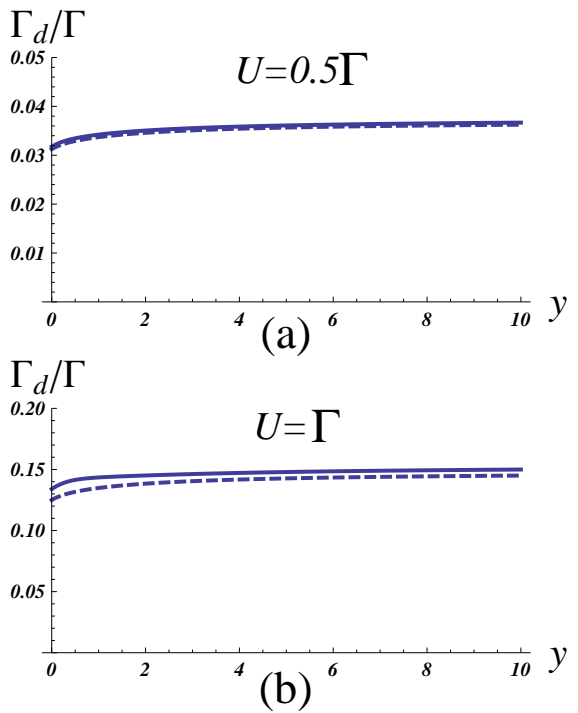}
\caption{(color online) Decoherence rate $\Gamma_d$, obtained from Eq.~(\ref{a14}), as a function of $y=\Gamma_2/\Gamma$ for $\Gamma_L=\Gamma_1=\Gamma$ and  $\Gamma_2=\Gamma_R=y\Gamma$  (solid lines), for two values of the interaction coupling $U$. Dashed lines show decoherence rate evaluated via the noise spectrum of the left SET, Eq.~(\ref{dec2}).} \label{fig5}
\end{figure}

Now we consider $\Gamma_2\not =0$. In this case the system of two separated SETs can be viewed as a detector monitoring the qubit's state (Fig.~\ref{fig4}a). We evaluate the decoherence rate $\Gamma_d$ as a function of $y=\Gamma_2/\Gamma$ by solving Eq.~(\ref{a14}) numerically. Consider first $\Gamma_L=\Gamma_1=\Gamma$ and $\Gamma_2=\Gamma_R=y\Gamma$, corresponding to maximal penetrability of each of the dots. We also assume that the energy levels of the dots are aligned, $E_1=E_2$. The results are shown by solid lines in Fig.~\ref{fig5} for two values of the interaction term: $U=0.5\Gamma$,  Fig.~\ref{fig5}(a) and $U=\Gamma$, Fig.~\ref{fig5}(b).

For comparison, we present $\Gamma_d$ as evaluated by Eq.~(\ref{dec2}) (dashed lines), where the correlator of charge inside the first SET (the left-hand one in Fig.~\ref{fig3}), $S_Q(0)$, has been calculated by neglecting the Coulomb interaction with the qubit. In fact, in the presence of the second SET, the Coulomb interaction does affect $S_Q$ by displacing the energy level of the left-hand SET by $U$ when the qubit's upper dot is occupied. However, the corresponding effect on the decoherence rate is on the order of $(U/\Gamma )^4$, which is within the limits of accuracy of Eq.~(\ref{dec2}). Figure~\ref{fig5} shows that $\Gamma_d$ is well reproduced by the fluctuation spectrum of the qubit's energy even for rather large values of $U$.

One finds from Fig.~\ref{fig5} that the decoherence rate in the case of measurement ($y>0$) exceeds that for the no-measurement case ($y=0$).
In fact, from a comparison between Fig.~\ref{fig4} and Fig.~\ref{fig5} one could conclude that the decoherence rate increases with the signal $\delta I$ for $0<y\lesssim 1$. However, for $y\gtrsim 1$ the signal drops with $y$, whereas $\Gamma_d$ reaches saturation. Nevertheless, Fig.~\ref{fig5} seems to support the common assumption that measurement generates decoherence \cite{gisin}.

\begin{figure}[h]
\includegraphics[width=9.5cm]{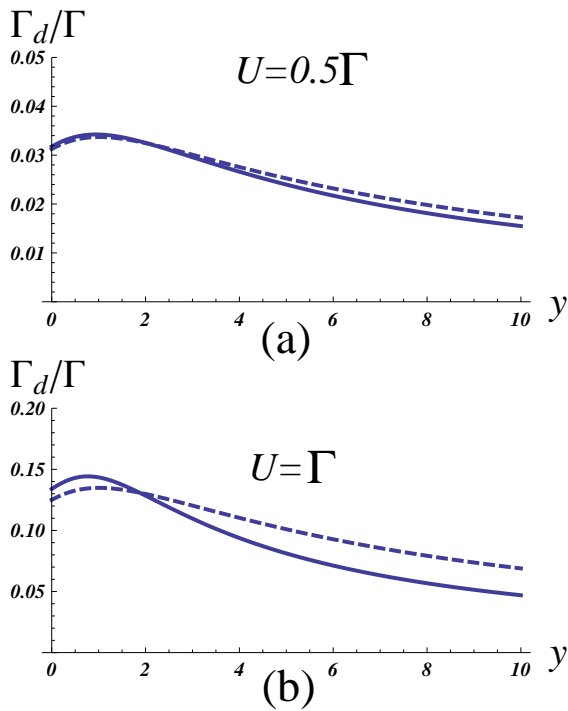}
\caption{(color online) Decoherence rate $\Gamma_d$ as a function of $y=\Gamma_2/\Gamma$ for $\Gamma_L=\Gamma_1=\Gamma_R=\Gamma$, $\Gamma_2=y\Gamma$ (solid line). Dashed lines show decoherence rate evaluated via the noise spectrum of the left-hand SET, Eq.~(\ref{dec2}).} \label{fig6}
\end{figure}

Consider now an asymmetric right-hand SET in Fig.~\ref{fig3}, by taking $\Gamma_L=\Gamma_1=\Gamma_R=\Gamma$ and $\Gamma_2=y\Gamma$. The results for the decoherence rate are shown in Fig.~\ref{fig6}. Here again the solid lines show exact evaluation of decoherence rate, given by Eq.~(\ref{a14}), whereas the dashed lines show an approximate evaluation of $\Gamma_d$ through Eq.~(\ref{dec2}). In contrast with Fig.~\ref{fig5}, the results look very surprising: They contradict the picture that measurement always destroys quantum coherence. Indeed, we find that by increasing the second SET's asymmetry, the decoherence rate {\em decreases\/} for $y\gtrsim 1$. Although the signal, $\delta I$, is also decreasing in this region (Fig.~\ref{fig4}), the total current still carries information on the qubit's state. In particular, one finds that decoherence rate becomes the same as for  $y=0$ (no measurement) at $y\simeq 2$ . This implies that the qubit measurement does not increase its decoherence rate. Moreover, one finds from Fig.~\ref{fig6} that for a large enough $y$, the decoherence rate drops substantially below its initial value at $y=0$. This means that the measurement can even diminish decoherence generated by a pure environment.

\begin{figure}[h]
\includegraphics[width=8cm]{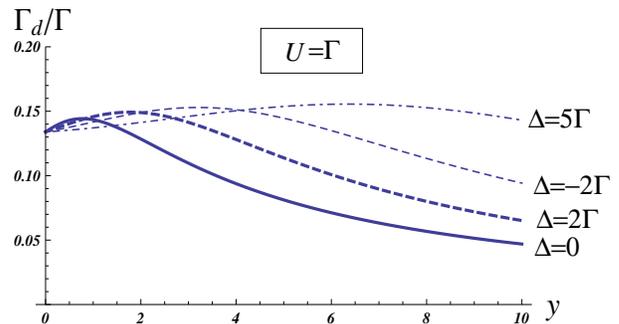}
\caption{(color online) Decoherence rate $\Gamma_d$ as a function of $y=\Gamma_2/\Gamma$ for different values of the two SETs misalignment, $\Delta$.} \label{fig8}
\end{figure}

Our results were obtained for aligned levels of the two SETs quantum dots, $\Delta =E_2-E_1=0$. The question is how the reduce of decoherence, shown in Fig.~\ref{fig6} is sensitive for this alignment. For this reason evaluated the decoherence rate as a function of $y$ for different values of $\Delta$ the results are presented in Fig.~\ref{fig8}. It follows from this figure that the phenomenon is rather stable with respect to the levels misalignment, until $\Delta$ is of the order of few $\Gamma$. There are other factors which can affect the reduction of decoherence by the second SET, for instance the environmental noise. Similarly, we anticipate that   if the level broading, generated by these factors is of the order of a few $\Gamma$, the phenomenon would survive.

In addition, our results presented in Figs.~\ref{fig4}--\ref{fig6}, indicate that there is no necessary relation between signal (information gain) and decoherence. Indeed, besides the region of $y\lesssim 1$, where the both quantities display similar behavior with $y$, this does not take place for $y\gtrsim 1$. For instance, one finds from  Figs.~\ref{fig4}(a) that the signal decreases to zero, when $y\to\infty$. However, in the same limit decoherence rate does not return to its value at $y=0$, corresponding to the no-signal regime. We can only conclude from Figs.~\ref{fig5}-\ref{fig6}, that decoherence is generated at most by a local environmental noise near the quantum system. A particular origin of this noise, like whether it is generated by measurement or not, in not relevant for decoherence. Therefore an understanding of a peculiar behavior of decoherence rate with the asymmetry of the second SET, can be achieved by analyzing the quantum noise.

\section{Interpretation}

Let us compare Fig.~\ref{fig5} with Fig.~\ref{fig6}. We want to understand the reason for the different behavior of the decoherence rate [or of the corresponding charge correlator, $S_Q(0)$, Eq.~(\ref{ac1})] in the two figures for $y\gtrsim 1$. Let us consider first the occupation probability of the left SET in the steady state, $\bar P_1$, shown in  Fig.~\ref{fig4}(b). The dashed line corresponds to the conditions in Fig.~\ref{fig5} and the solid line to those in Fig.~\ref{fig6}. In  both cases the occupation probability is saturated at large $y$. The solid line, however, ends up much closer to $1$ than the dashed line. That means that the amplitude of charge fluctuation is much smaller in the case of an asymmetric second SET, $\Gamma_2\gg\Gamma_R$, Fig.~\ref{fig6}, than for a symmetric one, $\Gamma_2=\Gamma_R$, Fig.~\ref{fig5}. As a result, the charge correlator drops with increasing asymmetry parameter $y$.

This dependence of the noise on the occupation of the SET's quantum dot can be seen from the following simple considerations. Consider the resonant current through the left-hand SET in Fig.~\ref{fig3} when it is decoupled from the right SET ($y=0$). The occupation in the steady state is given by $\bar P_1=\Gamma_L/(\Gamma_L+\Gamma_1)$. In the case of a symmetric left-hand SET, $\Gamma_L=\Gamma_1$, we find $\bar P_1=1/2$. The presence of the second SET ($y\not =0$) leads to $\bar P_1>1/2$. It looks as if the first SET were less strongly coupled to the reservoir, $\Gamma_1<\Gamma_L$. As a result, the charge correlator (\ref{ac1}) of the left dot, given by $S_0(0)=2\Gamma_L\Gamma_1/(\Gamma_L+\Gamma_1)^3$, drops. This demonstrates explicitly that an increase of the SET's occupation diminishes the corresponding charge correlator.

\begin{figure}[h]
\includegraphics[width=8.5cm]{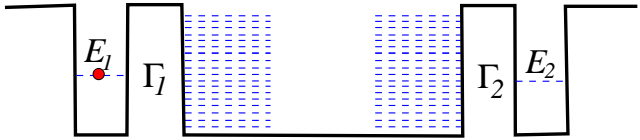}
\caption{(color online) An electron in two quantum wells separated by a reservoir.
}
\label{fig7}
\end{figure}

Now it remains to understand how a distant right-hand SET can increase the average occupation of the left-hand SET. This is an effect of quantum interference on large scales, which already appears in the motion of a single electron between two quantum wells separated by the reservoir (see Fig.~\ref{fig7}).
This problem has been recently investigated in Ref.~[\onlinecite{xin2}]. It was demonstrated there that in the case of aligned levels, $E_1=E_2$, an electron initially localized in the left-hand well can be found as $t\to\infty$ inside the the same well with probability $\Gamma_2^2/(\Gamma_1^{}+\Gamma_2^{})^2$. This implies that in the limit of $\Gamma_2\gg\Gamma_1$ the electron does not decay to the continuum, but remains locked in the left-hand well. This phenomenon is a result of quantum interference between two distant localized states.

Keeping in mind the increase in occupation of the left-hand SET generated by the distant right-hand SET in Fig.~\ref{fig7}, we can understand the difference in  occupation of the left-hand SET displayed in Fig.~\ref{fig4}(b) for the setup of Fig.~\ref{fig3}. In the case of $\Gamma_2\gg\Gamma_R$, the probability of penetration through the right-hand SET is suppressed. Therefore the right-hand SET can be considered as separated from the right lead. In this case it will resemble the setup of Fig.~\ref{fig7}, where the right well prevents decay of the left well to the middle reservoir for $\Gamma_2\gg\Gamma_1$, as explained in Ref.~[\onlinecite{xin2}] in great detail. If however $\Gamma_2 =\Gamma_R$, then the penetration of the right-hand SET is maximal. In this case it cannot be considered as detached from the right-hand reservoir. As a result the right-hand SET cannot hinder the decay of the electron from the left SET's quantum dot to the continuum, so that one cannot expect a substantial increase of occupation of the left SET. Using these arguments one can understand the behavior of $\bar P_1$ with $y$, shown in Fig.~\ref{fig4}(b), and therefore the surprising behavior of the decoherence rate, shown in Figs.~\ref{fig5} and~\ref{fig6}.

\section{Summary}

In this paper we have demonstrated how the retrieval of information from the environment of a quantum system affects the decoherence of the quantum system. For this purpose we considered an electrostatic qubit interacting with a fluctuating environment. The latter is represented by a current flowing through an SET near the qubit. The fluctuating charge inside the SET produces telegraph noise that affects the qubit's parameters. This generates decoherence, which causes information on the qubit's state to dissolve in the environment. For its retrieval, we introduce a distant SET that measures the environment. Solving the Schr\"odinger equation for the entire system and tracing out the environmental states, we evaluate the decoherence rate of the qubit in the presence of the second SET (measurement) or without it (no measurement).

In principle, as a result of decoherence, the information of the qubit's state is already lost in the environment before it is measured. Hence, one might assume that the actual measurement of the environment would not affect the decoherence rate of the qubit. Our analysis, however, shows that this is not the case: The actual retrieval of the qubit's information from the environment does affect the qubit's decoherence rate. Contrary to the common premise that any measurement can only increase decoherence, we find that the measurement can diminish the decoherence rate as well. This can be
interpreted as a result of destructive interference between the measuring device and the environment, leading to reduction of the local noise near the qubit.

Our analysis does not point to a general relation between the decoherence rate and the information gain. On the other hand, we find that the decoherence rate is determined by the local environmental noise, even in the case of indirect measurements.  This is true even beyond the weak coupling limit. In particular, a decrease of decoherence rate is always accompanied by a similar decrease of the local noise generated by the second SET. We suggested an interpretation of this effect as a manifestation of quantum interference on large scales, discussed in a previous publication.

We consider this work to be a first step in the investigation of indirect quantum measurements and decoherence. It would be very interesting to extend this research to different environments, where, for instance, the information regarding a quantum system is carried by emitted photons. We also expect that the effect of diminishing the decoherence (or the noise) with a distant detector, predicted in this paper, can be realized in various quantum systems and could then find various applications.

\begin{acknowledgements}We thank B. Svetitsky for
useful discussions and important suggestions to this paper.
One of us (S.G.) acknowledges the Department of Physics,
Beijing Normal University, and the State Key Laboratory for
Superlattices and Microstructures, Institute of Semiconductors, Chinese Academy of Sciences for supporting his visit.
This work was supported by the Israel Science Foundation under grant No.\ 711091, and NNSF of China under grants  No.\ 101202101 \& 10874176, and the Major State Basic Research Project of China under grants No. 2011CB808502 \& 2012CB932704.
\end{acknowledgements}


\begin{thebibliography}{999}
\bibitem{fn1} Note that Eqs.~(\ref{a4}) coincide with those obtained in Ref.~\onlinecite{g1}, except for the sign of the second (gain) term in the r.h.s. of Eq.~(\ref{a4e}) due to the Pauli factor ${\cal P}_2=-1$ of Eq.~(\ref{bp17}). This factor has not been taken into account in the earlier derivations of the quantum rate equations \cite{g1}.
\end{thebibliography}
\end{document}